\documentstyle[preprint,eqsecnum,aps]{revtex}
\tightenlines

\newfont{\lfig}{cmr10 scaled\magstep0} 
\begin{document}
\draft

%
%

\title{ Initial State Interactions for $K^-$-Proton Radiative Capture}
\author{Peter B. Siegel}
\address{California State Polytechnic University, Pomona CA 91768}
\author{Bijan Saghai}
\address{
Service de Physique Nucl\'{e}aire, CEA/DSM/DAPNIA, \\
Centre d'\'{E}tudes de Saclay, F-91191 Gif--sur--Yvette, France}

\bigskip

\bigskip

\date{\today}
\maketitle
\begin{abstract}
The effects of the initial state interactions on the $K^--p$
radiative capture branching ratios are examined and found to be
quite sizable.  A general coupled-channel
formalism for both strong and electromagnetic
channels using a particle basis is presented, and applied
to all the low energy $K^--p$ data with the exception of
the {\it 1s} atomic level shift.  Satisfactory fits are
obtained using vertex coupling constants for the electromagnetic
channels that are close to their expected SU(3) values.
\end{abstract}

\bigskip


\newpage

\section{\bf Introduction}
\par

The $K^-$-proton interaction is a strong multichannel
process [1], with the $\Lambda (1405)$ resonance
just below the $K^--p$ threshold at 1432 MeV.
At low energies, the $K^-$ can elastically
scatter off the proton, charge exchange to $\overline{K^0}-n$,
or scatter to ($\Sigma^+ \pi^-$), ($\Sigma^0 \pi^0$),
($\Sigma^- \pi^+$) or ($\Lambda \pi^0$) final states.
In addition, the electromagnetic radiative capture processes $K^- p \rightarrow
\Lambda \gamma (\Sigma^0 \gamma)$ are also possible [2].
Besides, the amplitudes of these latter reactions can be related to those of
the associated strangeness photoproduction, i.e. $\gamma p \rightarrow
 K^+ \Lambda (K^+ \Sigma^0)$, by the crossing symmetry.
Much effort has been done experimentally and theoretically
to understand this system.  In
particular, experiments to measure the branching ratios
of the radiative capture reactions $K^-p \rightarrow \Lambda \gamma$
and $K^-p \rightarrow \Sigma^0 \gamma$ were recently
performed [3] to help clarify the details of the reaction mechanism,
with a special interest in
the nature of the $\Lambda (1405)$ resonance [4]. However,
on one hand the result for the
 $\Lambda \gamma$ channel is
unexpectably smaller than both the previous measured value [5]
and those obtained through phenomenological models [2], and on the
other hand the measured branching ratio for the $\Sigma^0 \gamma$ final states
comes out significantly higher than the one for the other channel,
producing yet another mystery to this already complicated problem.

The most recent measurements of the threshold branching ratios with
stopped kaons, done
at Brookhaven [3] are:


  $$ R_{\Lambda \gamma} = {{\Gamma(K^-p \rightarrow \Lambda \gamma)} \over
         {\Gamma(K^-p \rightarrow  all)}} = .86 \pm .07 \pm .09 \times
10^{-3}~, $$

\noindent and

$$
  R_{\Sigma \gamma} = {{\Gamma(K^-p \rightarrow \Sigma^0 \gamma)} \over
            {\Gamma(K^-p \rightarrow  all)}} = 1.44 \pm .20 \pm .11 \times
10^{-3}.
$$

\noindent Existing calculations [2] overestimate
$R_{\Lambda \gamma}$ by a factor of 3 or 4 (except a few recent
phenomenological
 analysis [6,7] of the kaon photoproduction processes). The
pioneer calculations considered the $\Lambda (1405)$ in two different ways:
as an {\it s}-channel resonance [8] or as a quasi-bound ($K-N,\Sigma \pi$)
state [9-11]. In the quark-model approaches, this
hyperon is considered
as a pure $q^3$ state [12,13], a quasi-bound $\overline{K}$N state [14,15]
 or still as a hybrid ($q^3$+$q^4$$\overline{q}$...) state [16-18].
A number of potential model fits to the scattering data incorporate the
$\Lambda (1405)$ as a quasi-bound ($K-N,\Sigma \pi$) resonance [19-21].  It
would be
interesting to see if the radiative capture channels
can also be understood within a single model. Specially, since
the radiative capture data were taken to help distinguish between these two
possibilities.

     The diversified nature of the low energy data challenges theoretical
models.  Even if the analysis is restricted to the hadronic sector,
difficulties arise when trying to understand the $K^-p$ {\it 1s} atomic level
shift.  The sign of the $K^-p$ scattering length, extracted {}from this
experiment, is opposite to that
determined {}from K-matrix and potential model fits to the other hadronic
data.  This conflict poses interesting questions which are discussed
in Ref.[19,20,22].  Only one potential, Ref. [20], has been published which
is compatible
with all the low energy hadronic data.  We will examine the initial state
interactions for the radiative capture branching ratio for this potential.

     Calculations which focus on the radiative capture branching ratios
usually do not include the initial state interactions.
Only one group [15], which
uses the cloudy bag quark model, has included these in the electromagnetic
branching ratio calculation.  Since the interactions are
strongly coupled among the various channels, any meaningful
comparison with the data needs to include channel couplings.  We find here that
the effect of the initial state interactions is far {}from being negligible.
One limitation with this latter calculation is that no comparison is made with
the strong branching ratio data.  The other threshold branching ratios are
[23,24]:

$$
    \gamma = {{\Gamma(K^-p \rightarrow \pi^+ \Sigma^-)} \over
       {\Gamma(K^-p \rightarrow \pi^- \Sigma^+)}} = 2.36 \pm .04 ,$$

\noindent

$$
    R_c = {{\Gamma(K^-p \rightarrow charged ~particles)} \over
       {\Gamma(K^-p \rightarrow all)}} = .664 \pm .011 ,$$

\noindent and

$$
    R_n = {{\Gamma(K^-p \rightarrow \pi^0 \Lambda)} \over
       {\Gamma(K^-p \rightarrow ~all~ neutral~ states)}} = .189 \pm .015 .$$

\noindent They put tight constraints on the threshold amplitudes and potential
coupling strengths [17,20,21].  In fact, the five branching ratios are
amongst the most precise data in the strangeness sector. However, at present
there is no comprehensive analysis which includes both the hadronic and
electromagnetic branching ratios of the $K^-p$ system.

     The aim of this paper is to examine all the low energy data within a
single model and determine if it can be understood using
known coupling strengths and minimal SU(3) symmetry breaking for
relevant vertices in the
electromagnetic channels.
In doing so, we focus on how the two radiative capture
branching ratios are affected by the initial state interactions
among the different channels.  In order to unravel the essential
physics {}from the many channel system, in section 2 we will first
set up a general procedure to separate the strong (or initial state)
interactions {}from the electromagnetic ones.  As shown in section 2,
the initial state hadronic interactions can be described with 6 complex
numbers.  Thus we need a model of the interaction between the strong channels
to produce these 6 numbers.  In section 3 we examine two phenomenological
potential models, each of which fit the low-energy scattering data, the
resonance at 1405 MeV, and the hadronic branching ratios at threshold.  In one
potential, present work, the relative potential strengths between the various
channels are
guided by SU(3) symmetry.  For this potential, the parameters are adjusted to
fit
all the low energy $K^-p$ data with the exception of the {\it 1s} atomic level
shift
value of the $K^-p$ scattering length.  The other potential is {}from Ref.[20],
in
which the scattering length is compatible with the atomic level shift data.

     To clarify our discussion, we wish to underline here the nature of
the fitting parameters in the potential guided by SU(3) symmetry.  For the
hadronic channels, the relative potential strengths are given by a value
determined {}from SU(3) symmetry times a "breaking factor", which is equal
to 1, if the the relative channel couplings are SU(3)
symmetric.  This SU(3) structure is motivated by chiral symmetry [14].
For a good
fit to the low energy data we need to vary the relative strengths somewhat,
allowing the breaking factor to deviate {}from 1.  The final values of this
factor have no obvious physical significance.  The potential enables one to
estimate the effects of the initial state interactions {}from a potential which
gives a good fit to the low-energy data. In the electromagnetic channels,
 the radiative capture amplitudes are
derived {}from first order "Born" photoproduction processes, which
involve the meson-baryon-baryon coupling constants, $g_{Kp \Lambda}$,
$g_{Kp \Sigma}$, $g_{\pi \Sigma \Sigma}$, and $g_{\pi \Sigma \lambda}$.
These coupling
constants are related by SU(3) symmetry to the well known $\pi NN$ coupling
constant.  For the fit, these coupling constants are allowed to vary up to $\pm
50 \%$ {}from their SU(3) values.  Here the final values of these 4 parameters
will have physical significance and can be compared to values derived {}from
other
analyses. Thus we will examine if the radiative capture branching ratio data
can be understood
using vertex coupling constants for the electromagnetic channels that are
close to their expected SU(3) values when initial state interactions are
included {}from a hadronic interaction which fits the low energy hadronic data.

\vspace{.2in}

\section{\bf General Formalism}

        Consider a coupled-channel system consisting of {\it n} hadronic
channels
and one electromagnetic channel.  We assume that the interaction
between channels can be represented for each partial wave $l$
by real potentials of the form  $V_{ij}^l(\sqrt{s},k_i,k_j)$
where $\sqrt{s}$ is the
total energy and $k_i$ is the
momentum of channel {\it i} in the center-of-mass frame.
We will use the notation where the indices {\it i,j} are integers for the
strong channels and is $\gamma $ for the electromagnetic channels.
We will also assume that the transition matrix element for each
partial wave {}from
channel {\it i} to channel {\it j} can be derived {}from a coupled-channel
Lippmann-Schwinger equation:

$$
   T_{ij}(\sqrt{s},k_i,k_j) = V_{ij}(\sqrt{s},k_i,k_j) +
     \sum\limits_{m} \int V_{im}(\sqrt{s},k_i,q) G_m(\sqrt{s},q)
                     T_{mj}(\sqrt{s},q,k_j) q^2 dq,    \eqno(1) $$

\noindent where $G_i(\sqrt{s},q)$ is the propagator for channel {\it i}.
We suppress the index $l$, since for our problem only the $l=0$
partial wave is of interest.
The electromagnetic coupling is weak, and to a very good
approximation we can neglect the back coupling of the photon
channels.  Thus the T-matrix for radiative capture can be
written as:

$$
     T_{i \gamma}(\sqrt{s},k_i,k_{\gamma}) =
                V_{i \gamma}(\sqrt{s},k_i,k_{\gamma}) +
     \sum\limits_{m\ne \gamma} \int V_{im}(\sqrt{s},k_i,q) G_m(\sqrt{s},q)
                     T_{m \gamma}(\sqrt{s},q,k_{\gamma}) q^2 dq  .  \eqno(2) $$

Note that in the above equation there is no integration over the
photon's momentum.  There is only an integration over the
hadronic momentum $k_i$ in the $V_{i \gamma}$ potential.  This
means that only half off-shell information is needed for the
hadron-photon potential.  Since $\sqrt{s}$ and $k_\gamma$ are fixed
in the integral,
we can write $V_{i \gamma}(\sqrt{s},q,k_{\gamma})$
as:

$$
    V_{i \gamma}(\sqrt{s},q,k_\gamma) = {{V_{i \gamma}(\sqrt{s},q,k_\gamma)}
                 \over {V_{i \gamma}(\sqrt{s},k_i,k_\gamma)}}
                     V_{i \gamma}(\sqrt{s},k_i,k_\gamma), $$

or

$$
   V_{i \gamma}(\sqrt{s},q,k_\gamma) = v_{i \gamma}(q)
                                 V_{i \gamma}(\sqrt{s},k_i,k_\gamma).
\eqno(3)$$

\noindent Substituting this form for $V_{i \gamma}$ into equation (2) we obtain
for the T-matrix:

$$
    T_{i \gamma}(\sqrt(s),k_i,k_\gamma) = \sum \limits_{m \ne \gamma}
             M_{im}(\sqrt{s}) V_{m \gamma}(\sqrt{s},k_m,k_\gamma), $$

\noindent with the matrix $M_{im}$ defined as

\begin{eqnarray*}
   M_{im} &\equiv&  \delta_{im} +  \int V_{i,m}(\sqrt{s},k_i,q)
               G_m(\sqrt{s},q) v_{m \gamma}(q) q^2 dq \\
        & &  + \sum_{n \ne \gamma} \int \int V_{in}(\sqrt{s},k_i,q')
           G_n(\sqrt{s},q') V_{nm}(\sqrt{s},q',q) G_m(\sqrt{s},q)
           v_{m \gamma}(q) q'^2 dq' q^2 dq
             +  \cdots  .
\end{eqnarray*}

\noindent The state "m" is the last hadronic state before the photon is
produced.  Since all the on-shell momenta are determined {}from $\sqrt{s}$ we
have

$$
     T_{i \gamma}(\sqrt{s}) = \sum \limits_{m \ne \gamma} M_{im}(\sqrt{s})
                                              V_{m \gamma}(\sqrt{s}).
\eqno(4)$$

\noindent This form for the transition matrix to the photon channels is very
convenient, since it separates out the strong part {}from
the electromagnetic part of the interaction.  The matrix {\it M} is
determined entirely {}from the hadronic interactions and vertices.
In the absence of channel-coupling M is the unit matrix.
Any deviation {}from unity is related to the initial state interactions.
Note that {\it no assumptions were made on the form of the propagator
or the potentials connecting the hadronic channels}.  They
need not be separable.

Labeling the $K^--p$ channel as number 5, and defining $A_m(\sqrt{s})$
as $M_{5m}(\sqrt{s})$ we can write the scattering amplitude to the
photon channels as:

$$
    F_{K^-p \rightarrow \Lambda \gamma (\Sigma^0 \gamma)} =
             \sum A_m(\sqrt{s})
   f_{m \rightarrow \Lambda \gamma (\Sigma^0 \gamma)}.  \eqno(5)$$

\noindent where the $f_m$'s are the amplitudes to go {}from the
hadronic channel {\it m} to the appropriate photon channel.  These
amplitudes are derivable {}from diagrams representing the photoproduction
process.  The quantities $A_m$ are unitless complex numbers,
and contain all the information about the initial state
interactions for radiative capture.  Generally the sum over m is restricted to
states which have charged hadrons.  For the $K^--p$ process
the problem is greatly simplified, since there are only 3
channels which have charged hadrons: $\pi^+ \Sigma^-$,
$\pi^- \Sigma^+$ and $K^-p$.  To a very good approximation (see
section 3),
the $A_m$'s are the same for both the $\Lambda \gamma$ and
the $\Sigma^0 \gamma$ channels.  Thus, three complex numbers,
determined {}from the hadronic interactions, describe all the initial
state interactions for decays to both $\Lambda \gamma$ and
$\Sigma^0 \gamma$ final states.

     The result of Eq. (5) is essentially Watson's Theorem [25]
using a particle basis.  Watson's Theorem, which
also relates information about the strong interaction to that
of the electromagnetic process, uses an isospin basis.
The photoproduction amplitude is shown to have a phase
equal to the hadronic phase shift for a given isospin.  Eq. (5)
reduces to this result if there is only one strong channel.
In this case, $A$ is proportional to
$e^{i \delta}$ where $\delta$ is the phase-shift of
the strong channel.  For pion-nucleon photoproduction
it is useful to use an isospin basis since the T-matrix is diagonal and
both the photoproduction amplitude and
the hadronic phase shift can be determined {}from experiment.
It is especially useful if one isospin dominates (i.e. the $P_{33}$).
However, the T-matrix (or potential) for $K-N$, $\Sigma \pi$, $\Lambda \pi$
system is not diagonal in an isospin basis.  Watson's Theorem would
apply to the eigenphases of the coupled $K-N$, $\Sigma \pi$ system
for I=0, and the coupled $K-N$, $\Sigma \pi$, $\Lambda \pi$ system
for I=1.  Since these phases are not easily determined {}from experiment
the results of Watson's Theorem are not as useful in this case.  Also, in
the next section we point out that isospin breaking effects are
very important at low $K^--p$ energies.  Thus, in analyzing
threshold branching ratios, a particle basis is necessary.  Another
advantage of using Eq. (5) is that the interference of the "Born Amplitudes"
$f$ due to the initial state interactions of the hadrons is made
transparent.

\vspace{.2in}

\section{\bf Results and Discussion}

\subsection{The potentials for the strong channels}

        The two parts in determining the photoproduction rates in Eq. (4)
are the $A_m$, which are determined {}from the strong part of the
interaction, and the channel amplitudes $f_{m \rightarrow \Lambda \gamma
(\Sigma \gamma)}$.   For notation,
we will label the channels 1-8 as $\pi^+ \Sigma^-$,
$\pi^0 \Sigma^0$, $\pi^- \Sigma^+$, $\pi^0 \Lambda$,
$K^- p$, $\overline{K^0} n$, $\Lambda \gamma$, and $\Sigma^0 \gamma$
respectively.   We begin by discussing the determination of the $A_m$.
These were obtained
by using a separable potential and fitting to the available
low energy data on the strong channels.  We took $v_{i \gamma}$ in Eq. 3 to be
equal to $v_i$ in Eq. 6 below.  Two different separable
potentials were used: one guided by $SU(3)$ symmetry for the relative channel
couplings which fits
all the low energy data except the {\it 1s} atomic-level shift,
and one {}from Ref. [20] which fits all the low energy data including
the sign of the scattering length {}from the {\it 1s} atomic-level shift.
Values for
the $A_m$ at the $K^-p$ threshold for each fit are
listed in Table I.

\bigskip

{\lfig
\noindent {\lfig\bf TABLE I.} The $A_i$ values {}from Eq. (5) for two different
strong potentials.  The potential with approximate SU(3) symmetry fits
all low energy hadronic data except the {\it 1s} $K^-p$ atomic level shift.
The potential of Ref. [20] fits the atomic level shift as well.
}%

\medskip

\begin{center}
\begin{tabular}{|ccc|}\hline
$A_i$      &   Potential with Approximate   &   Potential of  \\
           &   $SU(3)$  Symmetry            & Tanaka and Suzuki [22] \\
\hline
$A_1$      &  (1.20, 0.52)                  & (1.49, -0.28)     \\
$A_2$      &  (-1.02, -.14)                 & (-1.10, 0.52)     \\
$A_3$      &  (0.83, -.23)                  & (0.71, -0.75)     \\
$A_4$      &  (-0.16, -0.34)                & (-0.30, -0.39)     \\
$A_5$      &  (-0.15, 1.06)                  & (2.01, 2.55)     \\
$A_6$      &  (1.18, -0.41)                 & (2.08, -1.12)     \\
\hline
\end{tabular}
\end{center}

\bigskip

     Following Ref. [21] the separable potentials for the strong
channels are taken to be of the form:

$$
   V^I_{ij}(k,k') = {{g^2} \over {4 \pi}} C^I_{ij} b^I_{ij}
                         v_i(k) v_j(k'),  \eqno(6)$$

\noindent where the $C^I_{ij}$ are determined {}from $SU(3)$ symmetry.  The
$b^I_{ij}$ are "breaking parameters" which are allowed to vary slightly
{}from unity.
The $v_i(k)$ are form factors, taken for this analysis to
be equal to $\alpha_i^2 / (\alpha_i^2 + k^2)$, and {\it g} is an overall
strength constant.  These potentials
are used in a coupled-channel Lippmann-Schwinger equation
with a non-relativistic propagator to solve for the cross-sections
to the various channels.  The data used in the fit are {}from Refs. [26-30].
The resonance at an energy of 1405 MeV was also fitted.
As discussed in Ref. [21] it was not possible to fit all the low-energy
data using potentials that had $b^I_{ij}=1$ for all i and j.  To get an
acceptable fit without including
the radiative capture data, it is
necessary to vary the $b^I_{ij}$ by at least $\pm 15 \%$ {}from unity.  To get
a very good fit to all
the data and determine the range of
the $A_m$,
we let the $b^I_{ij}$ vary {}from $0.5$ to $1.5$.  In Table II we list the
values we used for
the $I=0$ and $I=1$ potentials for our "best fit".  This "best fit" also
included the~~radiative capture data, and is discussed in

\bigskip

{\lfig
\noindent {\lfig\bf TABLE II.} The "best fit" values of $C^I_{ij}(b^I_{ij})$
for the potential of Eq. (6).
}%

\medskip

\begin{center}
\begin{tabular}{|cccc|}\hline
$C^{I=0}_{ij}$  &    $\Sigma \pi$      &    $KN$  &   \\ \hline
$\Sigma \pi$  &   $-2$ (0.50)        &   $-{{\sqrt{6}} \over 4}$ (1.29)& \\
$KN$          & $-{{\sqrt{6}} \over 4}$ (1.29) &  $-{3 \over 2}$ (1.43)&  \\
\hline

$C^{I=1}_{ij}$  &  $\Sigma \pi$      &   $\Lambda \pi$   &   $KN$    \\
\hline
$\Sigma \pi$  & $-1$ (0.50)     &   0   &  $-{1 \over 2}$ (1.37)  \\
$\Lambda \pi$ &     0        &   0   &  $ {\sqrt{6} \over 4}$     \\
$KN$ & $-{1 \over 2}$ (1.37) & $ {\sqrt{6} \over 4}$  & $-{1 \over 2}$
(0.50) \\ \hline

$\alpha_{\Sigma \i} = 974$ & $\alpha_{\Lambda \pi} = 886$ & $\alpha_{KN} =
 445 $  & $g^2 = 1.19 fm^2$ \\ \hline
\end{tabular}
\end{center}

%
%
%

\bigskip

\noindent the next section.  The elements are listed as
a product of the $C^{I}_{ij}$ values {}from SU(3) times $b^I_{i}$ which was
allowed to vary {}from $0.5$ to $1.5$.  Also listed
are the values for $\alpha_i$ in MeV/c and the overall strength $g^2$ {}from
Eq. (6).
We note that for this fit the $\Lambda (1405)$ is produced as a
$K-N (\Sigma \pi)$ bound state resonance [21] (see Fig. 4).

     The $A_i$ are a measure of how much the initial state interactions
enhance the single scattering amplitude.  Not all the $A_i$ are needed
in the $K^-p$ radiative decay calculation, since only channels which
have charged particles contribute.  Thus only $A_1$, $A_3$, and $A_5$
enter the calculation.  Also due to isospin symmetry in the $\Sigma \pi$
sector $A_1$, $A_2$, and $A_3$ must satisfy the relation: $A_1 + A_3 = -2A_2$.
In the absence of initial state interactions, $A_1=A_3=0$ and $A_5=1$.
As can be seen in Table I,
the magnitude of $A_1$, $A_3$, and $A_5$ are between 0.8 and 1.3.
Since $F_{K^-p \rightarrow \Lambda \gamma} = A_1 f_{\Sigma^- \pi^+
\rightarrow \Lambda \gamma} + A_3 f_{\Sigma^+ \pi^-
\rightarrow \Lambda \gamma} + A_5 f_{K^-p \rightarrow \Lambda \gamma}$,
cancellations amongst the various amplitudes can make the radiative
capture probability very sensitive to the initial state interactions.
Unfortunately, the $A_i$ cannot be directly determined
experimentally, and will have some model dependencies.  We tried
to estimate the model dependency for the potential of Eq. 6
by allowing the $b^I_{ij}$ to vary different amounts between 0.5 and 1.5 and
see how much the $A_i$
changed.  For acceptable fits to the data, excluding the atomic
{\it 1s} level shift, the $A_i$ varied only $\pm 20 \%$ in magnitude.

     An important aspect of the problem is to include the appropriate
isospin breaking effects due to the mass differences of the particles.
This was done as described in Ref. [21] by using the correct relativistic
momenta and reduced energies in the propagator.  The effects are very
important in calculating the threshold branching ratios, since the masses
of $\overline{K^0}-n$ are 7 MeV greater than the masses of $K^-p$.
As shown in Ref.[31], the Coulomb potential can be neglected when
calculating the branching ratios.  In Fig. 1 we plot the branching
ratios as a function kaon laboratory momentum $P_{Lab}$.
The three different curves for each ratio
correspond to different types of SU(3) breaking to be discussed later.
Note that the energy dependence is particularly
strong for branching ratios $\gamma$, $R_n$, $R_{\Lambda \gamma}$ and
$R_{\Sigma \gamma}$.

     For $\gamma$, which is the ratio of $\Sigma^- \pi^+$
production to $\Sigma^+ \pi^-$ the energy dependence is
easy to understand.  For a model as the one presented here which does not
include the $\Lambda (1405)$ as an {\it s}-channel resonance,
the reaction $K^-p \rightarrow \pi^+ \Sigma^-$ cannot occur in a single
step.  This is a double charge exchange reaction, and needs to undergo two
single charge exchange steps with the middle one being neutral.  {}From the
total cross-section data (See Fig. 2), the most important neutral channel in
low-energy $K^-p$ scattering comes out to be $\overline{K^0} n$.
This causes the ratio $\gamma$ to have a strong energy
dependence near the $\overline{K^0} n$ threshold.   Since the $\overline{K^0}
n$
channel is also
important in $\Lambda \pi$ production, the ratio $R_n$ also varies rapidly
with energy near threshold.  Thus for an
accurate comparison with the data, one needs to use a particle basis
in calculating the photoproduction branching ratios at threshold.
We note that if an {\it s}-channel resonance was the dominating process
in the $\Sigma \pi$ reaction, then the
ratio $\gamma$ would not have as rapid an energy
dependence near the $K^-n$ threshold.  It is also interesting that
$\Gamma (K^-p \rightarrow \Lambda \gamma)$ is substantially less than
$\Gamma (K^-p \rightarrow \Sigma^0 \gamma)$ at energies below the
$\overline{K^0} n$
threshold and
greater at energies above.
Experimental data of these branching ratios
near the $K^-p$ threshold would help clarify the nature of
the $\Lambda (1405)$.

%
%

\bigskip

\subsection{The electromagnetic channels}

        We now turn our attention to the most important part of the
calculation, the amplitudes for the $\Lambda \gamma$ and $\Sigma^0 \gamma$
channels.  As discussed previously,
only the amplitudes for the three charged channels
will contribute to the radiative capture amplitude in Eq. (5):
$K^-p \rightarrow \Lambda \gamma (\Sigma^0
\gamma)$, $\Sigma^{\pm} \pi^{\mp} \rightarrow \Lambda \gamma (\Sigma^0
\gamma)$.  Here we will use
the amplitudes obtained {}from the diagrams shown in Fig. 3.  These
diagrams are the leading order contributions to photoproduction [32].
We include the most important amplitudes which are the
"extended Born terms", including the $\Lambda$ and
the $\Sigma^0$ exchange terms, and the vector meson exchange terms
($K^*$, $\rho$).  The expressions for these terms and their relative
importance are given in Appendix I.

%
%
%
%

\bigskip

     There are four coupling constants which enter in the photoproduction
amplitudes: $g_{KN \Lambda}$, $g_{KN \Sigma}$, $g_{\pi \Sigma \Sigma}$,
and $g_{\pi \Sigma \Lambda}$.  Since there are only two branching
ratios to fit, we need to limit the method of our search.  We
investigated three cases: {\it a}) assume exact SU(3) symmetry for the coupling
constants with
$g_{\pi NN} = 13.4$ and vary the F-D mixing ratio $\alpha$ to best
fit the data, {\it b)} assume $\alpha = 0.644$, $g_{\pi NN} = 13.4$
and vary the coupling
constants slightly {}from their SU(3) values for a best fit, and
{\it c}) use the $A_m$ {}from the potential of Ref. [20] and
SU(3) symmetry for the coupling constants to see if a fit
of the radiative decay branching ratios is possible.  For the search, we
weighted each data point equally, and thus the two radiative capture branching
ratios did not have a great affect on the hadronic parameters.

     In the first case, we assumed exact SU(3) symmetry with
$g_{\pi NN} = 13.4$ and varied the F-D mixing ratio $\alpha$ for
a best fit to the data.  We found an acceptable fit with a $\chi^2$ per
data point of 2.47 for $\alpha = 1.0$.  The branching ratios for
this fit are $ \gamma$ = 2.25, $ R_c$ = 0.66 and $ R_n$ = 0.17
 for the strong channels, and
$ R_{\Lambda \gamma}$ =$1.22 \times 10^{-3}$ and $ R_{\Sigma \gamma}$
=$1.47 \times 10^{-3}$ for the electromagnetic
channels.
 Although this is not the accepted value for $\alpha$,
it is remarkable to get a fit with only one
adjustable variable.

     In the next case, we fix $\alpha$ to be 0.644.  The search is done
using MINUIT code [33] on
13 parameters: the three ranges for the strong channels, the six breaking
factors for the strong channels $b^I_{ij}$, $g_{Kp \Lambda}$,
$g_{Kp \Sigma}$, $g_{\pi \Sigma \Sigma}$, and $g_{\pi \Sigma \lambda}$.  We
allowed the $b^I_{ij}$ to vary $\pm 50 \%$, $\pm 40 \%$, and $\pm 30 \%$
{}from unity while the 4 coupling constants $g_{Kp \Lambda}$,
$g_{Kp \Sigma}$, $g_{\pi \Sigma \Sigma}$, and $g_{\pi \Sigma \lambda}$
varied by $\pm 50 \%$, $\pm 40 \%$, and $\pm 30 \%$
{}from their SU(3) values respectively.  The range parameters $\alpha$ were
allowed to vary {}from 200 to 1000 MeV/c.  The results for the branching
ratios and the reduced $\chi^2$ are listed in Table III.  The first
column lists the percentage that the parameters, except $g_{Kp \Lambda}$,
were allowed to vary {}from their SU(3) values (or in the case of the
$b^I_{ij}$'s {}from unity).  The
second column lists the percentage that $g_{Kp \Lambda}$ was allowed to vary
{}from its SU(3) value of -13.2.  We also tried to find a satisfactory fit in
which $g_{Kp \Lambda}$ was as close to -13.2 as possible.  A fit was found in
which
$g_{Kp \Lambda}$ was varied only $\pm 20 \%$, while the other parameters were
allowed to vary $\pm 50 \%$.  The first row of Table III shows these results.
We call this our "best fit" since the most well determined coupling constants,
$g_{KN \Lambda}$ and $g_{KN \Sigma}$ are close to their SU(3) values, with
$g_{KN \Sigma}$ only $50 \%$ high.  Our best fit values for the coupling
constants are
$g_{KN \Lambda}=-10.6$, $g_{KN \Sigma}=5.8$, $g_{\pi \Sigma \Sigma}=-7.2$,
and $g_{\pi \Sigma \Lambda}=-5.0$. Notice that our values for the two first
coupling constants are in agreement with those obtained {}from strangeness
photoproduction [7,34] and hadronic sector [35,36] analyses.
The electromagnetic branching ratios change drastically if the initial
state interactions are excluded {}from the calculation.
We obtain $R_{\Lambda \gamma}=.56 \times 10^{-3}$ and
$R_{\Sigma \gamma} = .12 \times 10^{-3}$ without the initial state
interactions. The two branching ratios are hence decreased by roughly
a factor of 2 and more than one order of magnitude, respectively,
by switching off the initial state interactions.

Graphs of the different
fits for the five branching ratios and total cross sections
as a function of kaon laboratory momentum are shown in Fig. 1 and 2,
respectively.  In Fig. 4 we plot the $\Sigma \pi$ spectrum normalized
to the data of Hemmingway [37].  As in Ref.[14], we plot
$k^\pi_{c.m.} |T_{\Sigma \pi \rightarrow \Sigma \pi}|^2$, where
$T_{\Sigma \pi \rightarrow \Sigma \pi}$ is the T-matrix in the I=0
sector for $\Sigma \pi \rightarrow \Sigma \pi$ scattering.
In each figure, the solid line corresponds to the "best fit"
parameters, the dotted line to $\pm 40 \%$ SU(3) breaking for all
the parameters, and the dashed line to
$\pm 30 \%$ breaking for all the parameters. In each case the
$\Lambda (1405)$ is produced as a bound state resonance as in Ref [21].

     The $K^-p$ scattering length obtained {}from our best fit is
(-.63 + .76 i) fm.  This compares closely with the value {}from Ref.
[32] of (-.66 + .64 i) fm.  These values, however,
have the opposite sign for the real part {}from
that extracted {}from the {\it 1s} $K^-p$ atomic level shift data [38].
Since the atomic level shift data is still puzzling [39], we did not try
to fit it in our search.  This discrepancy has been discussed in
detail in Ref. [20] with some interesting results.  Hence,
we used the $A_m$ obtained {}from the potential
of Ref. [20] which fit all the hadronic low energy data and has the same
sign for the scattering length as the atomic level shift data.  We were
able to reproduce their results using their non-relativistic potentials.
{}From Table I we see that $A_1$ and $A_3$ do not differ too much {}from
those obtained with the "SU(3) guided" potential.
However, $A_5$ is much different
in magnitude and its real part has the opposite sign.  Perhaps this
is because the atomic {\it 1s} shift and hence the scattering length has
the opposite sign.
For the potential of [20] the resulting
radiative capture branching ratios using coupling constants {}from
SU(3) symmetry
are $R_{\Lambda \gamma} =
17.5 \times 10^{-3}$ and $R_{\Sigma \gamma} = 3.29 \times 10^{-3}$,
which are far {}from the experimental values.  For satisfactory
agreement with the radiative capture branching ratios, the
coupling constants would have to deviate {}from their SU(3) values
by an unreasonable amount.
The reason for the bad agreement is that
$A_5$ is very large and its real part is positive.  In order to
obtain a small value for $\Lambda \gamma$ production, the amplitudes
have to cancel in Eq. (5).  Since the relative
signs of the $f_{m \rightarrow \Lambda \gamma (\Sigma^0 \gamma)}$ are
fixed by SU(3) symmetry, the $A_m(\sqrt{s})$ have to have appropriate
relative phases to cause this cancellation.  The
$A_m$ {}from the potential guided by SU(3) symmetry have this feature.
\bigskip

{\lfig
\noindent {\lfig\bf TABLE III.} Branching ratios and $\chi^2$ per data point
for
different amounts of SU(3) breaking.  Column 2 lists the variation in the
coupling constant $g_{Kp \Lambda}$.  Column 1 lists the variation in the
other parameters.
}%

\medskip

{\small
\begin{center}
\begin{tabular}{|cccccccc|}\hline
All except $g_{Kp \Lambda}$ & $g_{Kp \Lambda}$ & $\chi^2/N$ & $\gamma$ & $R_c$
& $R_n$ & $R_{\Lambda \gamma} \times 10^{3}$ &
         $R_{\Sigma \gamma} \times 10^{3}$ \\ \hline
$\pm 50 \%$ & $\pm 20 \%$ & 1.76 & 2.31 & .661 & .164 & 1.09 & 1.55 \\
$\pm 50 \%$ & $\pm 50 \%$ & 1.21 & 2.35 & .659 & .194 & 0.89 & 1.46 \\
$\pm 40 \%$ & $\pm 40 \%$ & 1.54 & 2.32 & .659 & .179 & 1.04 & 1.53 \\
$\pm 30 \%$ & $\pm 30 \%$ & 2.94 & 2.20 & .652 & .174 & 1.31 & 1.65 \\  \hline
\hline
Experiment  &             &      & 2.36 $\pm$ .04 & .664 $\pm$ .011 & .189
$\pm$ .015 &
                   .86 $\pm$.07  & 1.44 $\pm$ .20 \\  \hline
\end{tabular}
\end{center}
}

\bigskip

\section{\bf Conclusions}

     We have done a comprehensive analysis of all the low energy
data, except the {\it 1s} atomic level shift, on the $K^-p$ system.  To
facilitate the analysis, we derived an expression for the radiative
capture cross section which separates out the strong interaction {}from
the electromagnetic ones.  The initial state interactions can be
described by six complex amplitudes, $A_m$, with only three of them
relevant to the radiative capture process.
For the strong part of the interaction
we choose a separable potential whose relative potential strengths
were guided by SU(3) symmetry.  This potential is phenomenological and
serves to produce appropriate $A_m$ {}from the low energy scattering and
resonance data.  The radiative capture amplitudes
are derived {}from first order "Born" photoproduction processes, and
are determined {}from meson-baryon-baryon coupling constants, whose values
are related by SU(3) symmetry to the well known $\pi NN$ coupling constant.

     We found a number of good fits in which the coupling constants were
close to their expected SU(3) values.  For these fits, the relative
coupling strengths in the strong channels were guided by $SU(3)$ symmetry.
In all of the fits, the $\Lambda (1405)$
is produced as a bound $K-N (\Sigma \pi)$ resonance, and the initial state
interactions were very important for the radiative capture branching ratios.
The ratio $R_{\Lambda \gamma}$ varies roughly by a factor of 2, and the ratio
$R_{\Sigma \gamma}$ by more than a factor of 10 due to the initial
state interactions.

        Results presented in this paper, reproduce well enough the
existing strong and electromagnetic data {}from threshold up to $P_{K}^{\rm
lab}
\approx$ 200 MeV/c. Our predictions, specially for the branching ratios, show
clearly the need for more experimental investigations ; one of
the main motivations being to clarify the nature of the $\Lambda (1405)$
resonance. Such measurements are
planned at DA$\Phi$NE [40] using the tagged low energy kaon
beam and may also be achieved at Brookhaven and KEK.

\section{\bf Acknowledgements}

        We would like to thank J.C. David, C. Fayard, G.H. Lamot, Andreas
Steiner
and Wolfram Weise for many helpful discussions and suggestions
regarding this
work.  We are grateful to the Institute for Nuclear Theory (Seattle), for
an stimulating and pleasant stay, where the idea of this collaboration
emerged. One of us (PS) would like to thank the Centre d'Etudes
de Saclay for the hospitality extended to him.

\newpage

\begin{center}
\Large {\bf Appendix I}\\

\end{center}

\bigskip

     In this Appendix we will summarize the contributions to the
photoproduction amplitudes shown in Fig. 3.  Here we will
write the expressions for the $K^- p \rightarrow \Lambda \gamma$
amplitude.  The amplitudes for the $K^- p \rightarrow \Sigma^0
\gamma$, $\Sigma^{\pm} \pi^{\mp} \rightarrow \Lambda \gamma$,
and $\Sigma^{\pm} \pi^{\mp} \rightarrow \Sigma^0 \gamma$ processes
will be the same with appropriate masses and coupling constants.

     To lowest order the amplitude for the $K^-p \rightarrow \Lambda
\gamma$ reaction f is the sum of three amplitudes:

\medskip

$$
      f_{K^-p \rightarrow \Lambda \gamma} = F_{Born} + F_\Sigma + F_{K^*},
$$

\medskip

\noindent which correspond to the
the Born, the $\Sigma^0$, and the $K^*$ diagrams shown in Fig 3.

     The Born amplitude is derived in Ref. [32] and is given by:

\medskip

$$
 F_{Born} = - \sqrt{{E_{\Lambda} + m_{\Lambda}} \over {2 m_{\Lambda}}}
             {{g_{Kp \Lambda} e} \over {2 m_p}}
            (1 + {{k_{\gamma}} \over {E_{\Lambda}+m_{\Lambda}}}
              (1 + \kappa_p + \kappa_{\Lambda})),
$$

\medskip

\noindent at the $K^-p$ threshold.  The "$\Sigma$" term
is also derived in Ref. [32] and is given by:

\medskip

$$
 F_{\Sigma} = - \sqrt{{E_{\Lambda} + m_{\Lambda}} \over {2 m_{\Lambda}}}
             {{g_{Kp \Sigma} e} \over {2 m_p}}
             \kappa_{\Sigma \Lambda} {{\sqrt{s}-m_\Lambda} \over
                      {\sqrt{s} + m_\Sigma}}.
$$

\medskip

\noindent In a similar manner, the $K^*$ exchange term
can be evaluated.  In this case, there is a vector and a
tensor piece.  We obtain for the amplitude:

\medskip

$$
 F_{K^*} = - \sqrt{{E_{\Lambda} + m_{\Lambda}} \over {2 m_{\Lambda}}}
            [ {{g^V_{K^*p \Lambda} e} \over {2 m_p}}
           {{\kappa_{K^*K} k^2_\gamma (\sqrt{s}-m_p)} \over
                      {(t - m^2_{K^*})(E_\Lambda + m_\Lambda)}}
          +  {{g^T_{K^*p \Lambda} e} \over {2 m_p}}
           {{\kappa_{K^*K} k^2_\gamma (\sqrt{s}-m_p)} \over
                      {(t - m^2_{K^*})2 m_p}}
              ({{m_\Lambda + m_p} \over {m_\Lambda + E_\Lambda}})].
$$

\medskip

In the absence of initial state interactions, the differential cross section is
given by:

\medskip

$$
 {{d \sigma} \over {d \Omega}} = {{(E_\Lambda + m_\Lambda)
               (E_p + m_p)} \over {64 \pi^2 s}}
                   {{P_\gamma} \over {P_K}} |f_{K^-p \rightarrow
                           \Lambda \gamma}|^2.
$$

\medskip

     The first part of $F_{Born}$ has the largest magnitude.  The other
pieces are reduced by kinematical factors, with $F_{K^*}$ giving
the smallest contribution.  The $K^*$ exchange makes up about
$2 \%$ of the amplitude.  Thus, the uncertainty in the vector
and tensor coupling constants are not so important, and we fixed
them to be the SU(3) values.  Also, since the calculation is
not particularly sensitive to the values of the electromagnetic
couplings, we fixed them at their excepted values.  Values for
the constants which were held fixed during the search are
listed in Table IV.

\newpage

{\lfig
\noindent {\lfig\bf TABLE IV.} Values of the coupling constants which were held
constant.
}%

\medskip

\begin{center}
\begin{tabular}{|ll|}\hline
$\kappa_p = 1.793$  & $g^V_{K^*p \Lambda} = -4.5 $  \\
$\kappa_\Lambda = -.613$  & $g^T_{k^*p \Lambda} = -16.6 $  \\
$\kappa_{\Sigma \Lambda} = 1.6$  & $g^V_{K^*p \Sigma} = -2.6 $   \\
$\kappa_{K^*K} = 1.58$   &  $g^T_{K^*p \Sigma} = 3.2 $   \\
$\kappa_{\rho \pi} = 1.41$  & $g^V_{\rho \Sigma \Lambda} = 0$  \\
$\kappa_{\Sigma^-} = -2.157$    & $g^T_{\rho \Sigma \Lambda} = 11.1$  \\
$\kappa_{\Sigma^+} = 1.42$    & $g^V_{\rho \Sigma \Sigma} = -5.2$ \\
$\kappa_{\Sigma^0} = .619$    & $g^T_{\rho \Sigma \Sigma} = -12.8$  \\
\hline
\end{tabular}
\end{center}

\bigskip

     The search was done on the more important coupling constants,
$g_{K p \Lambda}$, $g_{K p \Sigma}$, $g_{\pi \Sigma \Sigma}$, and
$g_{\pi \Sigma \Lambda}$.  At the $K^-p$ threshold, using the
coupling constants of Table III, the radiative capture amplitudes
are:

\medskip

$$
     f_{K^-p \rightarrow \Lambda \gamma} = {e \over {2 m_p}}
              (-g_{Kp \Lambda}(1.28) - g_{Kp \Sigma}(.2) -.75),
$$

$$
    f_{K^-p \rightarrow \Sigma^0 \gamma} = {e \over {2 m_p}}
              (-g_{Kp \Sigma}(1.32) - g_{Kp \Lambda}(.15) + .02),
$$

$$  f_{\pi^+ \Sigma^- \rightarrow \Lambda \gamma} = {e \over {2 m_\Sigma}}
    (-g_{\pi \Sigma \Lambda}(.835) + g_{\pi \Sigma \Sigma}(.19) + .21),
$$

$$
   f_{\pi^+ \Sigma^- \rightarrow \Sigma^0 \gamma} = {e \over {2 m_\Sigma}}
    (-g_{\pi \Sigma \Sigma}(.95) + g_{\pi \Sigma \Lambda}(.153) + .19),
$$

$$
  f_{\pi^- \Sigma^+ \rightarrow \Lambda \gamma} = {e \over {2 m_\Sigma}}
    (g_{\pi \Sigma \Lambda}(1.16) - g_{\pi \Sigma \Sigma}(.19) -.21),
$$

$$
  f_{\pi^- \Sigma^+ \rightarrow \Sigma^0 \gamma} = {e \over {2 m_\Sigma}}
   (-g_{\pi \Sigma \Sigma}(1.28) + g_{\pi \Sigma \Lambda}(.153) + .19),
$$

\medskip

\noindent where the three terms in parenthesis correspond to the three
amplitudes described above.  The above equations show the
relative importance of the different
contributions to radiative capture at the $K^-p$ threshold.  The best fit
values for these coupling constants are summarized in Table V.

\bigskip

{\lfig
\noindent {\lfig\bf TABLE V.} "Best fit" values of the coupling constants for
the electromagnetic amplitudes.
}%

\medskip

\begin{center}
\begin{tabular}{|ll|}\hline
$g_{Kp \Lambda} = -10.6$  & $g_{Kp \Sigma} = 5.8$  \\
$g_{\pi \Sigma \Sigma} = -7.2$  & $g_{\pi \Sigma \Lambda} = -5.0$  \\
\hline
\end{tabular}
\end{center}

\newpage

\begin{center}
\Large {\bf References}\\

\end{center}

\bigskip

\noindent ~[1] See, e.g., A.J.G. Hey and R.L. Kelly, {\it Phys.~Rep.~}
{\bf 96}, 71 (1983).

\noindent ~[2] See, e.g., J.~Lowe {\it Nuovo Cimento} {\bf 102A}, 167 (1989);
and references therein.

\noindent ~[3] D.A.~Whitehouse {\it et al.},
{\it Phys.~Rev.~Lett.~}{\bf 63}, 1352 (1989).

\noindent ~[4] See, e.g., R.C.~Barrett {\it Nuovo Cimento} {\bf 102A}, 179
(1989).

\noindent ~[5] J.~Lowe {\it et al.}, {\it Nucl.~Phys.~}
{\bf B209}, 16 (1982).

\noindent ~[6] R.~Williams, and C.~Ji,~S.~Cotanch,
{\it Phys.~Rev.~}{\bf C46}, 1617 (1992).

\noindent ~[7] J.C. David, C. Fayard, G.H. Lamot, F. Piron, and B. Saghai, in
{\it Proceedings

of the 8th Symposium on Polarization Phenomena in
Nuclear Physics}, Bloom-

ington, September 15-22, 1994, Ed. S. Vigdor
{\it et al.}, to appear;
J.C. David,

Ph.D. Thesis, Univerity of Lyon (1994), in French.

\noindent ~[8] T.A.~DeGrand~and~L.~Jaffe, {\it Ann.~Phys.~(N.Y.)~}
{\bf 100}, 425 (1976); T.A.~De-

{}~Grand. {\it ibid}, {\bf 101}, 496 (1976).

\noindent ~[9] G.~Rajasekaran, {\it Phys.~Rev.~}{\bf D5}, 610 (1972).

\noindent [10] R.H.~Dalitz, S.F.~Tuan, {\it Ann.~Phys.~}
{\bf 10}, 307 (1960).

\noindent [11] R.H.~Dalitz, T.-C.~Wong, and G.~Rajasekaran,
{\it Phys.~Rev.~}{\bf 153}, 1617 (1967).

\noindent [12] M.~Jones, R.H.~ Dalitz, R.R.~Horgan, {\it Nucl.~Phys.~}
{\bf B129}, 45 (1977).

\noindent [13] J.D.~Darewych, R.~Koniuk and N.~Isgur,
{\it Phys.~Rev.~}{\bf D32}, 1765 (1985).

\noindent [14] E.A.~Veit, B.K.~Jennings, A.W.~Thomas, and
R.C.~Barrett, {\it Phys.~Rev.~}{\bf D31},

{}~1033 (1985).

\noindent [15] Y.S.~Zhong, A.W.~Thomas, B.K.~Jennings, and
R.C.~Barrett, {\it Phys.~Rev.~}{\bf D38},

{}~837 (1988).

\noindent [16] E.~Kaxiras, E.J.~Moniz, and M.~Soyeur, {\it Phys.~Rev.~}{\bf
D32}, 695 (1985).

\noindent [17] G.~He and R.H.~Landau, {\it Phys.~Rev.~}
{\bf C48}, 3047 (1993).

\noindent [18] M.~Arima, S. Matsui, and K.~Shimizu, {\it Phys.~Rev.~}
{\bf C49}, 2831 (1994).

\noindent [19] J. Schnick and R.H. Landau, {\it Phys.~Rev.~Lett.~}{\bf 58},
1719 (1987).

\noindent [20] K.~Tanaka and A.~Suzuki, {\it Phys.~Rev.~}
{\bf C45}, 2068 (1992).

\noindent [21] P.B.~Siegel and W.~Weise,
{\it Phys.~Rev.~} {\bf C38}, 2221 (1988).

\noindent [22] K.S. Kumar and Y. Nogami,{\it Phys.~Rev.~}{\bf D21}, 1834
(1980).

\noindent [23] R.J.~Nowak {\it et al.}, {\it Nucl.~Phys.~}
{\bf B139}, 61 (1978).

\noindent [24] D.N.~Tovee {\it et al.}, {\it Nucl.~Phys.~}
{\bf B33}, 493 (1971).

\noindent [25] K.M.~Watson, {\it Phys.~Rev.~}{\bf 95},
228 (1954).

\noindent [26] J. Ciborowski {\it et al.}, {\it J. Phys.~}{\bf G 8}, 13 (1982).

\noindent [27] D. Evans {\it et al.}, {\it J. Phys.~}{\bf G 9}, 885 (1983).

\noindent [28] W.E. Humphrey and R.R. Ross, {\it Phys.~Rev.~}{\bf 127}, 1305
(1962).

\noindent [29] J.K.~Kim, Columbia University Report, Nevis
149 (1966).

\noindent [30] M. Sakitt {\it et al.}, {\it Phys.~Rev.}{\bf 139}, 719 (1965).

\noindent [31] P.B. Siegel, {\it Z. Phys.}{\bf A 328}, 239 (1987).

\noindent [32] R.L.~Workman and H.W.~Fearing, {\it Phys.~
Rev.~}{\bf D37}, 3117 (1988).

\noindent [33] F. James, in {\it Proceedings of the 1972 CERN Computing
and Data Processing

{}~School},Pertisan, Austria, 1972 (CERN Report 72-21);
F. James and M. Roos,

{}~{\it MINUIT Functional Minimization and Error
Analysis}, D506-Minuit, CERN

{}~(1989).

\noindent [34] R.A.~Adelseck and B.~Saghai, {\it Phys.~Rev.~}
{\bf C42}, 108 (1990).

\noindent [35] A.D. Martin,  {\it Nucl.~Phys.~} {\bf B179}, 33 (1981) ;
J. Antolin,  {\it Z.~Phys.~C} {\bf31}, 417

{}~(1986).

\noindent [36] M. Bozoian, J.C. van Doremalen, and H.J. Weber,
{\it Phys. Lett.~}{\bf 122B}, 138

{}~(1983).

\noindent [37] R. Hemingway, {\it Nucl. Phys.}{\bf B253}, 742 (1985).

\noindent [38] J.D.~Davies {\it et al.}, {\it Phys. Lett.~}{\bf 83B}, 55
(1979);
 M.~Izycki {\it et al.}, {\it Z. Phys.~}{\bf A297},

{}~11 (1980); P.M.~Bird,
A.S.~Clough, and K.R.~Parker,  {\it Nucl.~Phys.~}
{\bf A404}, 482

{}~(1983).

\noindent [39] C.J. Batty {\it Nuovo Cimento} {\bf 102A}, 255 (1989).

\noindent [40] See, e.g., G. Pancheri (Ed.), {\it Proceedings of the workshop
on Physics and

{}~Detectors for DA$\Phi$NE }, Frascati (Italy), April 9-12, 1991,
Servizio Documen-

{}~tazione dei Laboratori Nazionali di Frascati; FINUDA Collaboration,
{\it A De-

{}~tector for Nuclear Physics at DA$\Phi$NE }, Frascati Report LNF-93/021,
(1993).

\newpage

\centerline{\bf Figure Captions}

\bigskip

\noindent Figure 1.  The five branching ratios, defined in the text,
are plotted as a function
of $K^-$ laboratory momentum: {\it a}) branching ratio $\gamma$, {\it b})
$R_c$,
{\it c}) $R_n$, {\it d}) $R_{\Lambda \gamma}$, and {\it e}) $R_{\Sigma
\gamma}$.
The three curves correspond to different amounts of SU(3)
breaking as listed in Table III.  The solid curve corresponds to our
"best fit" parameters: the vertex couplings of Table V and the strong
couplings of Table II.  This corresponds to the first line in
Table III.  The dotted curve is for $\pm 40 \%$ variation in all the
parameters, the third line in Table III.  The dashed curve is for
$\pm 30 \%$ variation in all the parameters, the last line in Table III.
The data points at threshold [3,19,20] with error bars are also shown.
The $\overline{K^0}-n$ threshold is at 89.4 MeV/c.

\bigskip

\noindent Figure 2.    Cross sections are compared with the experimental
data for the six strong
channels and the two electromagnetic channels: {\it a}) $K^-p$ elastic
scattering,
{\it b}) $K^-p \rightarrow \overline{K^0} n$, {\it c})
$K^-p \rightarrow \pi^0 \Lambda$, {\it d})
$K^-p \rightarrow \pi^+ \Sigma^-$, {\it e}) $K^-p \rightarrow \pi^0 \Sigma^0$,
{\it f}) $K^-p \rightarrow \pi^- \Sigma^+$, g) $K^-p \rightarrow \Lambda
\gamma$,
and h) $K^-p \rightarrow \Sigma^0 \gamma$.
The three curves for each cross section
correspond to different amounts of SU(3) breaking as in Fig. 1.

\bigskip

\noindent Figure 3.  The main diagrams which contribute to
the radiative capture amplitude for
$K^-p \rightarrow \Lambda \gamma$:
{\it a), b), c)} and {\it f)} are the "Born" terms, {\it d)} the
"$\Sigma$" or cross term, and {\it e)} the vector meson $K^*$ exchange term.
The
$K^-p \rightarrow \Sigma^0 \gamma$ reaction and the $\Sigma \pi$
reactions $\pi^{\pm}  \Sigma^{\mp} \rightarrow \Lambda \gamma (\Sigma^0
\gamma)$
will have similar terms.

\bigskip

\noindent Figure 4.  The $\Sigma \pi$ mass spectrum normalized to the data of
Ref. [37] is plotted as a function of the $\Sigma \pi$ center of mass energy.
The three curves correspond to different amounts of SU(3) breaking as in Fig.
1.

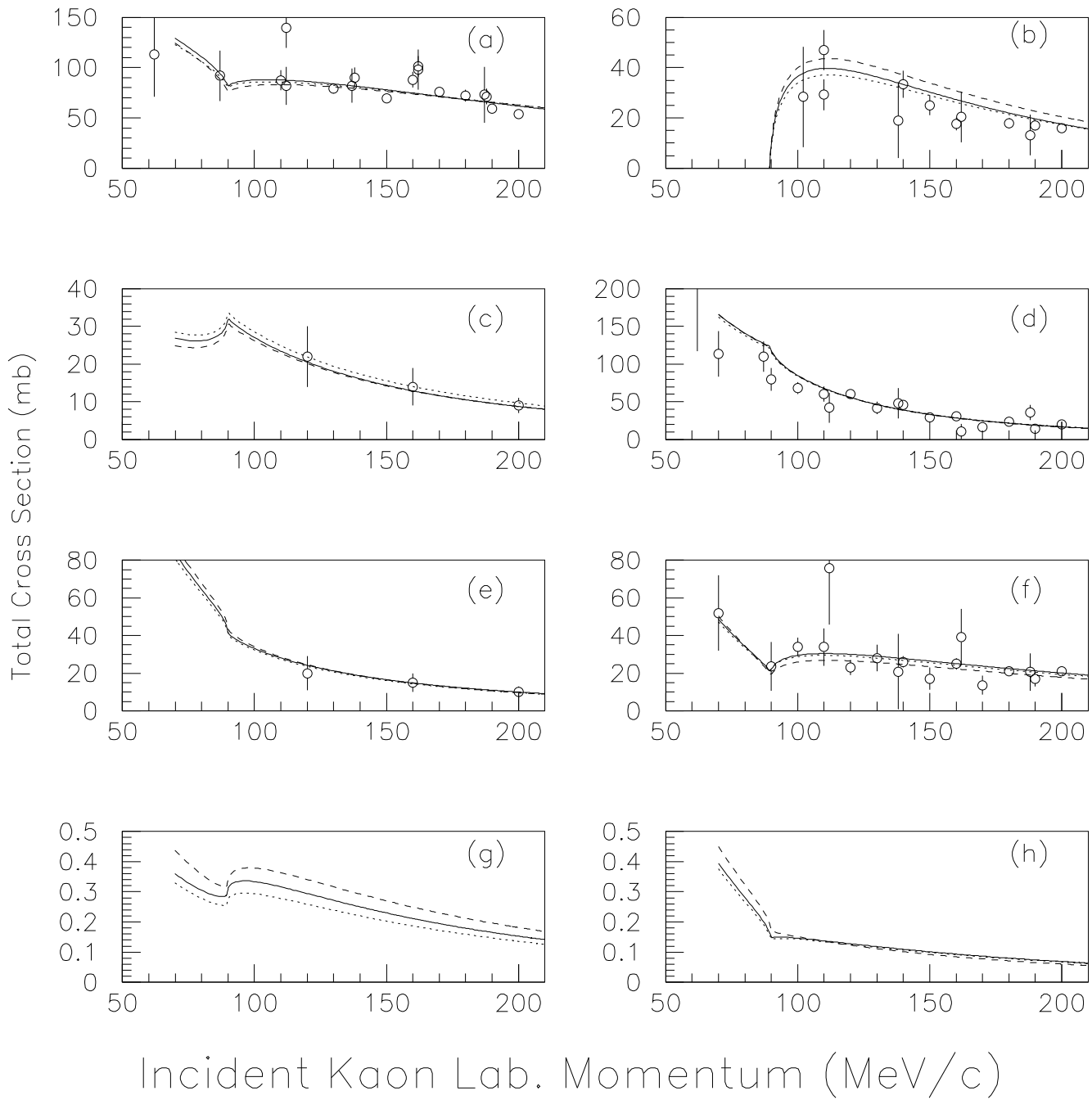
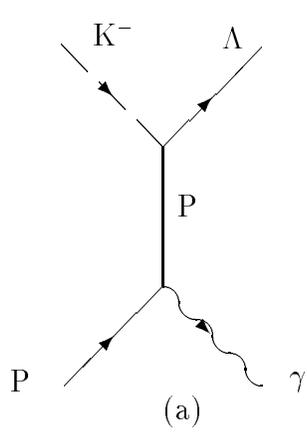
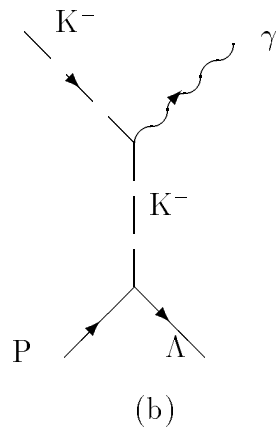
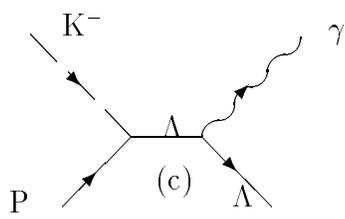
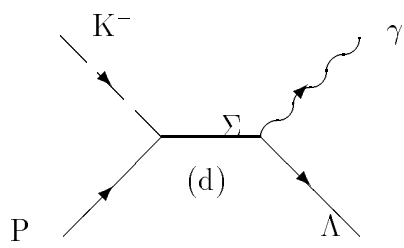
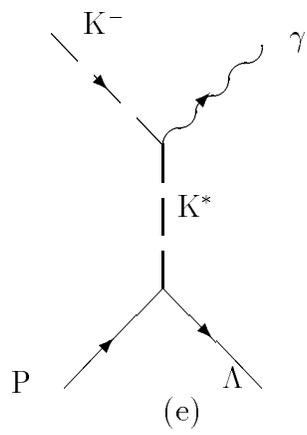
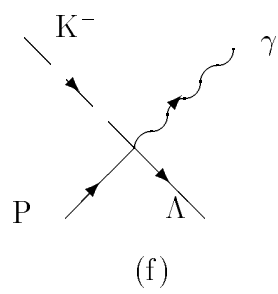
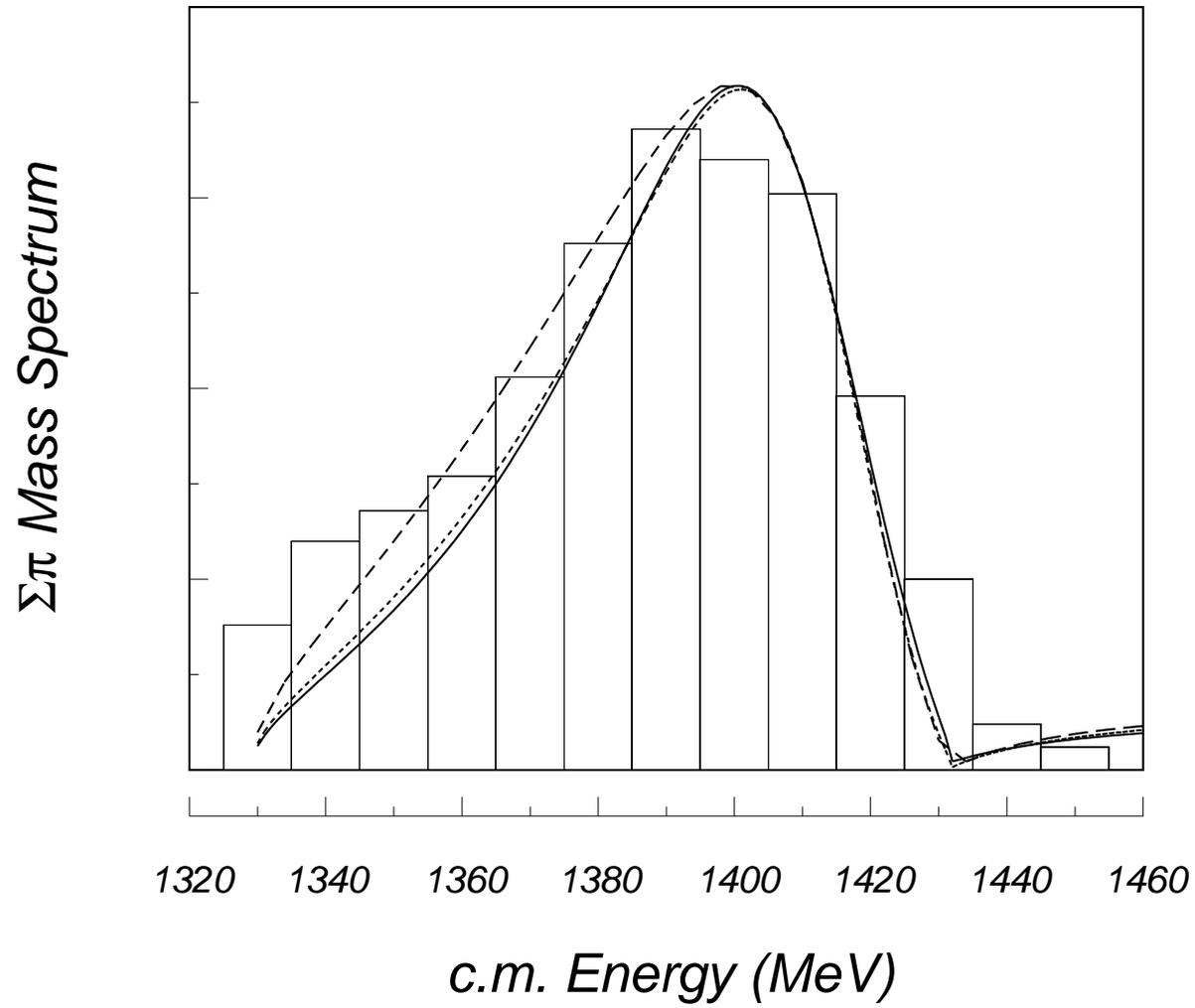
\end{document}